\documentclass[prl,superscriptaddress,twocolumn,nopreprintnumbers,
floatfix,nofootinbib]{revtex4}

\usepackage{graphicx,url,amssymb,amsmath,longtable,rotating,color,units,
wasysym,subfigure,epsfig,multirow}
\usepackage[dvipsnames]{xcolor}

\usepackage[plainpages=false, colorlinks=true, anchorcolor=blue, linkcolor=blue, citecolor=blue, bookmarks=false]{hyperref}

\begin{document}

\title{Gravitational Waves from Orphan Memory}
\author{Lucy O. McNeill}
\author{Eric Thrane}\email{eric.thrane@monash.edu}
\author{Paul D. Lasky}
\affiliation{Monash Centre for Astrophysics, School of Physics and Astronomy, Monash University, VIC 3800, Australia}


\begin{abstract}
Gravitational-wave memory manifests as a permanent distortion of an idealized gravitational-wave detector and arises generically from energetic astrophysical events.
For example, binary black hole mergers are expected to emit memory bursts a little more than an order of magnitude smaller in strain than the oscillatory parent waves.
We introduce the concept of ``orphan memory": gravitational-wave memory for which there is no detectable parent signal.
In particular, high-frequency gravitational-wave bursts ($\gtrsim$ kHz) produce orphan memory in the LIGO/Virgo band. 
We show that Advanced LIGO measurements can place stringent limits on the existence of high-frequency gravitational waves, effectively increasing the LIGO bandwidth by orders of magnitude.
We investigate the prospects for and implications of future searches for orphan memory.
\end{abstract}

\maketitle

The detection of gravitational waves by LIGO and Virgo \cite{GW150914} has opened up new possibilities for observing highly-energetic phenomena in the Universe.
It was recently shown that ensembles of binary black hole detections can be used to measure gravitational-wave memory \cite{Lasky2016}: a general relativistic effect, manifest as a permanent distortion of an idealized gravitational-wave detector \cite{Zeldovich1974, Braginsky1987, Christ1991, favata09, Favata2009c, favata11}.
It is not easy to detect memory.
The memory strain is significantly smaller than the oscillatory strain; $\sim20$ times smaller for GW150914.

For gravitational-wave bursts, the memory strain increases monotonically with a rise time comparable to the burst duration; e.g., $\tau\sim10$ ms for GW150914 \cite{Lasky2016}.
For sufficiently short bursts (with timescales that are short compared to the inverse frequency of the detector's sensitive band), the memory is well-approximated by a step function, or equivalently an amplitude spectral density proportional to $1/f$, where $f$ is the frequency.
It follows that the memory of a high-frequency burst introduces a significant low-frequency component which extends to frequencies arbitrarily below $1/\tau$.
If the parent burst is above the detector's observing band, this can lead to ``orphan memory": a memory signal for which there is no detectable parent.

There are a number of mechanisms that can lead to orphan memory.
In the example above, a high-frequency burst outside the observing band creates in-band memory.
This is the premise of memory burst searches in pulsar timing arrays \cite{Seto2009, Pshirkov10, Vanhaasteren2010, Nanograv2015, Wang2015}, which look for memory from merging supermassive black holes for which the oscillatory signal is out of band.
Orphan memory can also be sourced by phenomena other than gravitational waves, e.g., neutrinos \cite{epstein78, turner78}, although the probability of detection from known sources is small.
In principal, it is possible for beamed gravitational-wave sources to produce orphan memory signals when the oscillatory signal is beamed away from Earth.  
In practice, however, the number of orphan detections from beaming will be small compared to the number of oscillatory detections. 
In this Letter, we focus on memory where high-frequency gravitational-wave bursts produce orphan memory in LIGO/Virgo.

\begin{figure}[h!]
  \centering
    \includegraphics[width=0.5\textwidth]{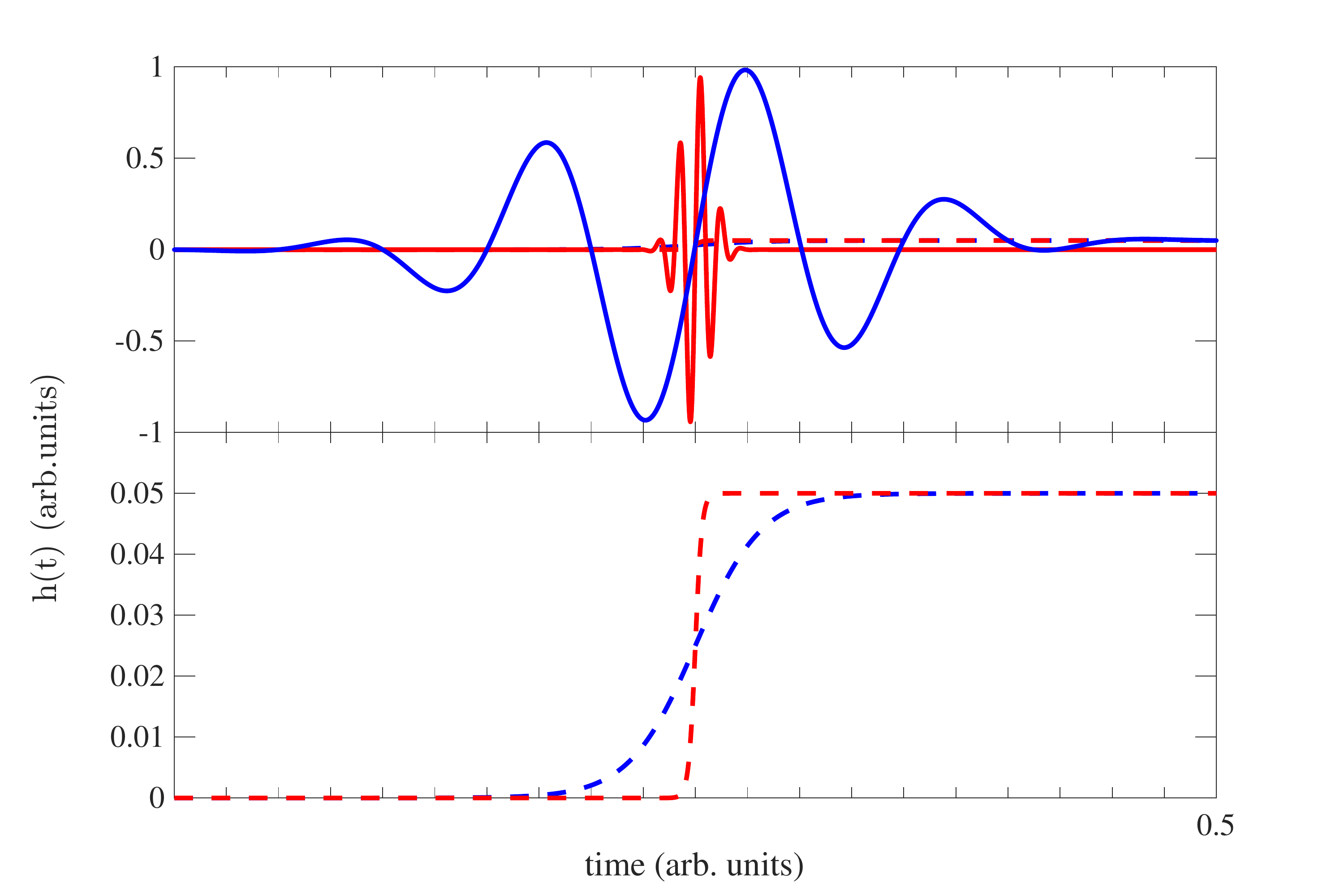}
  \caption{
Strain time series for a gravitational-wave burst.
The top panel shows both the burst (solid curves) and memory (dashed curves) strains for two bursts of the same amplitude.
The high-frequency burst (red) has frequency ten times the low-frequency burst (blue).
The bottom panel shows an enlarged version of the memory time series'.
As the frequency of the burst increases, the rise time approaches zero and the memory is well-approximated by a step function.
}
  \label{fig:waveforms}
\end{figure}

{\it Scaling relations.}
As a starting point it is useful to investigate scaling relations for gravitational-wave bursts.
For a gravitational-wave source with timescale, $\tau$, frequency $f_0\approx1/\tau$, and energy $E_{\rm gw}$, the strain amplitude scales as
\begin{align}
	h_0^{\rm osc}\sim\frac{E_{\rm gw}^{1/2}}{f_0^{1/2}d},\label{hosc}
\end{align}
where $d$ is the distance to the source, and throughout we use natural units, $c=G=1$.
A sine-Gaussian waveform is well described by these assumptions, and so we work with sine-Gaussian waveforms in the analysis that follows. 
In Figure~\ref{fig:waveforms} we show two sine-Gaussian bursts (top panel) with their corresponding memory waveforms approximated by $\tanh$ functions (bottom panel).

\begin{figure}[h!]
  \centering
    \includegraphics[width=0.5\textwidth]{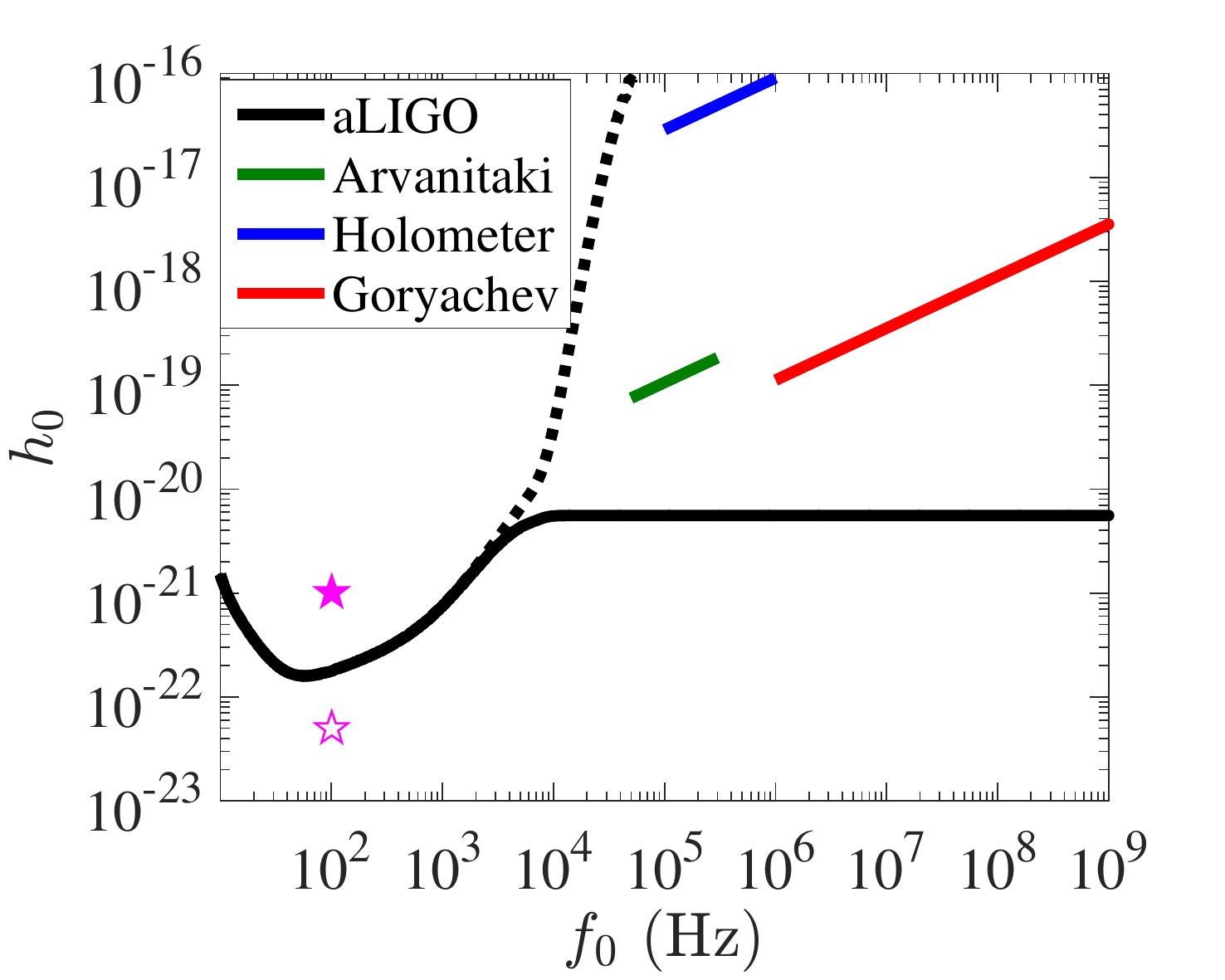}
  \caption{
  The dashed black curve shows the sine-Gaussian amplitude $h_0$, with frequency $f_0$, required for an average signal-to-noise-ratio of $\langle S/N \rangle=5$ in Advanced LIGO operating at design sensitivity.
  The solid black curve shows the same sensitivity to sine-Gaussian amplitude $h_0$ except that we include memory in the matched filter calculation, effectively extending the LIGO band to arbitrarily high frequencies.
  Around a few kHz, the memory becomes comparably important to the oscillatory signal.
  The memory strength is calculated using the fiducial value of $\kappa$; see Eq.~\ref{eq:kappahat}.
  The colored curves show the $h_0$ sensitivity of dedicated high-frequency gravitational-wave detectors: Arvanitaki \cite[green]{Arvanitaki}, Goryachev \cite[red]{Tobar2014}, Holometer \cite[blue]{Holometer2015}.
  For the fiducial value of $\kappa$, Advanced LIGO can detect memory bursts with higher significance than dedicated high-frequency detectors.
The filled star indicates the maximum strain of GW150914 and the frequency of peak emission~\cite{GW150914}.
The unfilled star indicates the expected memory from GW150914~\cite{Lasky2016}.
}
  \label{fig:h0f}
\end{figure}

On the other hand, the memory scales as \cite{Braginsky1987}
\begin{align}
	h_0^{\rm mem}\sim\frac{E_{\rm gw}}{d}.
\end{align}
At first glance it might appear that memory strain can exceed the oscillatory strain for sufficiently large $E_{\rm gw}$. 
However, the maximum gravitational-wave frequency (for ultra-relativistic systems) is given by $f_{\rm max}\sim 1/r_s \sim 1/M_s$, where $r_s, M_s$ are the Schwarzschild radius, mass.
In order to obtain a conservative estimate of the maximum possible memory, we assume that the entire remnant black hole mass is radiated away as gravitational waves.
Thus, the maximum memory occurs when $f_0=1/E_{\rm gw}$, implying the maximum possible memory is
\begin{align}
	h_{\rm max}^{\rm mem}\sim\frac{E_{\rm gw}^{1/2}}{f_0^{1/2}d},\label{hmemmax}
\end{align}
which scales like the oscillatory strain, Eq.~(\ref{hosc}).

While $h_{\rm max}^{\rm mem}$ and $h_0^{\rm osc}$ scale the same way, $h_{\rm max}^{\rm mem}$ is always smaller.
We define a memory efficiency factor 
\begin{align}
	\kappa\equiv h^{\rm mem}/h^{\rm osc}.
\end{align}
where $h^\text{mem}$ and $h^\text{osc}$ are respectively the measured memory and oscillatory strain in some detector.
The numerical value of $\kappa$ depends on the inclination and sky position of the source.
For an interferometeric detector such as LIGO, we can estimate the typical value of $\kappa$ by simulating an ensemble of binary black holes with random inclination angle and sky position, and using the memory waveforms of Ref. \cite{Favata2009a}.
We find
\begin{equation}\label{eq:kappahat}
 \hat\kappa\equiv
 \sqrt{\langle h_\text{mem}^2 \rangle / \langle h_\text{osc}^2 \rangle} = 0.044 .
\end{equation}
The angled brackets denote ensemble averages over angles.
This is not far from the estimated value~\cite{Lasky2016} for GW150914 $\kappa = 1/20$.

In this Letter, we assume a fiducial value of $\kappa=1/20$.
We stress that our fiducial $\kappa$ value, while plausible for efficient memory emission from highly relativistic objects such as binary black holes, leads to an overestimation of the memory signal from less relativistic systems.
For non-relativistic systems, $h_0^{\rm mem}$ can be very small compared to $h_0^{\rm osc}$.
If a gravitational-wave observatory were ever to measure $\kappa>1$, this would be surprising as it would seem to indicate a significant quantity of missing energy, perhaps due to beaming.

{\it High-frequency gravitational waves.}
At design sensitivity, the LIGO/Virgo detectors will operate between $\sim10-2000$ Hz \cite{noisecurve}.
Here, we consider high-frequency gravitational waves, which we define to be $\gtrsim2000$ Hz, and as high as $10^{15}$ Hz.
A number of astrophysical sources may emit high-frequency gravitational waves \cite[for a review, see Ref.~][]{Cruise2012}. 
These include black hole evaporation \cite[$10^{10}-10^{15}$ Hz;][]{Nakamura1997,Greene2012}, dark matter collapse in stars \cite[$\sim2$ GHz;][]{Kurita2015}, cosmic strings \cite{Damour1}, and Kaluza-Klein modes in higher-dimensional theories \cite{Seahra2005,Clarkson2007}.
Of these, the source that produces the loudest orphan memory is probably dark matter collapse in stars because black holes are maximally relativistic.
Such an event at $\unit[8]{kpc}$ should produce a memory signal with amplitude $h_0\approx 2\times10^{-25}$.
Given our fiducial value of $\kappa$, the average signal-to-noise ratio in Advanced LIGO is small: $\langle S/N \rangle \sim 10^{-3}$.

We estimate the sensitivity of Advanced LIGO to orphan memory from generic high-frequency sources by calculating the matched filter signal-to-noise ratio for our fiducial $\kappa=1/20$, high-frequency source.
The expectation value for the optimal matched filter signal-to-noise ratio is 
\begin{equation}
    \left<S/N\right>^2 = 4 \mathrm{Re} \int \frac{|\tilde{h}|^2}{S_h(f)} df \approx 4 \Bigg[ \frac{h_0^2}{S_h(f_0)} \frac{1}{f_0}\Bigg],
\end{equation} 
where the approximation holds for sine-Gaussians with frequency $f_0$ and amplitude $h_0$.


The dashed black curve in Figure~\ref{fig:h0f} shows the sine-Gaussian amplitude $h_0$ necessary for an average signal-to-noise ratio $\left<S/N\right>=5$ detection in Advanced LIGO operating at design sensitivity as a function of burst frequency $f_0$.
The solid black curve shows the same $h_0$ versus $f_0$ sensitivity curve except that we include memory in the matched filter calculation.
This has the effect of extending the LIGO observing band to sources for which the dominant oscillatory component has arbitrarily high frequencies.
For burst frequencies higher than a few kHz, the memory becomes more easily detectable than the oscillatory burst.
The colored curves show $h_0$ verus $f_0$ sensitivity for dedicated high-frequency detectors, which we discuss presently.

We compare the Advanced LIGO sensitivity curve to several dedicated high-frequency detectors.
Fermilab's ``Holometer" (labeled with a blue curve in Figure~\ref{fig:h0f}) is a pair of co-located $\sim\unit[40]{m}$, high-powered Michelson interferometers, sensitive to gravitational wave frequencies $\unit[10^5-10^6]{Hz}$.
It has reached an amplitude spectral density of $\unit[\sim7\times10^{-20}]{{Hz}^{-1/2}}$~\cite{Holometer2015}.
The Bulk Acoustic Wave (labeled with a red curve in Figure~\ref{fig:h0f}) cavity is a proposed resonant mass detector, sensitive to $\unit[10^6-10^9]{Hz}$ projected amplitude spectral density of $\sim\unit[10^{-22}]{Hz^{-1/2}}$~\cite{Tobar2014}.
The detector is labeled in plots as
``Goryachev" using the first author from~\cite{Tobar2014}.
The final proposed detector that we consider here consists of optically levitated sensors, sensitive to frequencies between $\unit[50-300]{kHz}$, with projected sensitivity to $\sim \unit[3\times10^{-22}]{Hz^{-1/2}}$~\cite{Arvanitaki}.
It is labeled in Figure~\ref{fig:h0f} with a green curve and denoted ``Arvanitaki" after the first author of~\cite{Arvanitaki}.

Comparing the Figure~\ref{fig:h0f} colored sensitivity curves for dedicated high-frequency detectors with the solid black sensitivity curve for Advanced LIGO, we see that---given our fiducial value of $\kappa$---Advanced LIGO will detect orphan memory before currently-proposed, dedicated, high-frequency detectors observe an astrophysical burst.

Two effects, not included in Figure~\ref{fig:h0f}, will tend to make it harder to detect high-frequency bursts compared to low-frequency memory detection.
First, high-frequency detectors produce false positives at a higher rate than Advanced LIGO.
Second, the memory search template bank is trivially small.
All orphan memory looks the same: like a step function.
In order to span the space of oscillatory bursts, it is likely that many more templates must be used.

This result has an interesting implication.
If high-frequency detectors observe a detection candidate, Advanced LIGO should look for a corresponding memory burst.
A coincident memory burst could provide powerful confirmation that the high-frequency burst is of astrophysical origin.
Similarly, if Advanced LIGO detects orphan memory, it may be worthwhile looking for coincident bursts in dedicated high-frequency detectors.

\begin{figure}[h!]
  \centering
  \includegraphics[width=0.5\textwidth]{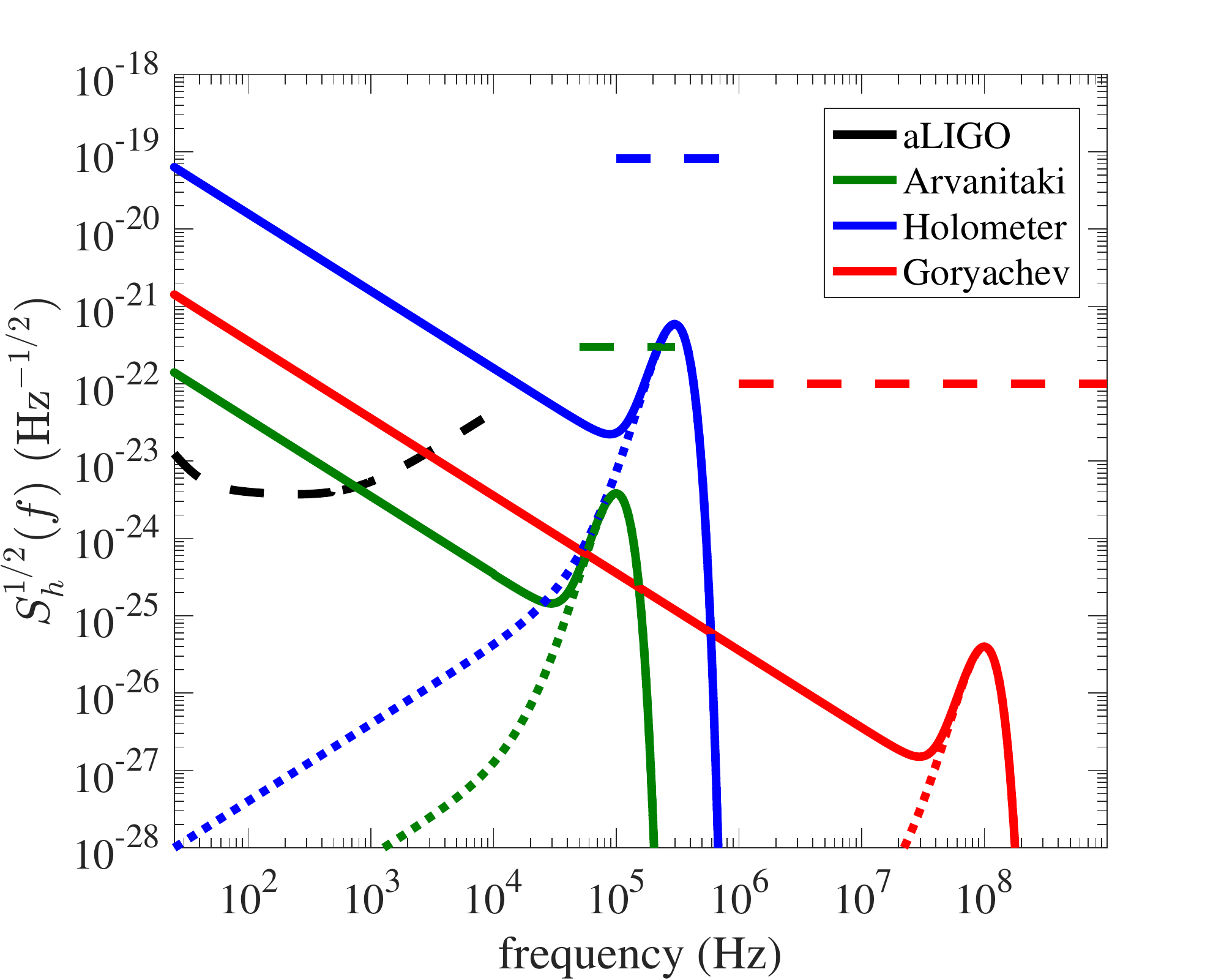}
  \caption{Strain amplitude spectral density.
  The dashed curves represent the noise in three different detectors: Advanced LIGO (black) and three dedicated high-frequency detectors (colored).
  For each dedicated detector, we plot the amplitude spectral density for a sine-Gaussian burst in the middle of the observing band (colored dotted peaks).
  The peak height is tuned so that the oscillatory burst can be observed with a signal-to-noise ratio $\langle S/N \rangle = 5$.
  The solid colored lines shows the amplitude spectral density when we include the memory calculated with our fiducial value of $\kappa$.
  The memory bursts produce large signals in Advanced LIGO, with $\langle S/N \rangle$ ranging from 300 to $10^5$.
  }
  \label{fig:ASD1}
\end{figure}

In Figure~\ref{fig:h0f}, we plot sensitivity curves in terms of $h_0$: the amplitude of a sine Gaussian burst.
It is also useful to frame our results in terms of amplitude spectral density $S_h(f)^{1/2}$.
In Figure~\ref{fig:ASD1}, we show the noise amplitude spectral densities for the three high-frequency detectors included in Figure~\ref{fig:h0f}, denoted with dashed red, blue, and green curves.
The dashed black curve shows the noise amplitude spectral density of Advanced LIGO.

We also plot the amplitude spectral density for three sine-Gaussian bursts with memory.
The frequency of each burst is matched to the observing bands of different high-frequency detectors.
The colors are chosen so that, e.g., the red burst spectrum matches with the red Goryachev detector.
The burst amplitude is tuned so that $\langle S/N \rangle=5$ in the associated high-frequency detector.
The solid curves show the memory + oscillatory component of the signal while the dotted curves show only the oscillatory component.

While the oscillatory matched filter signal-to-noise ratio is 5 in each high-frequency detector, the associated LIGO memory signal-to-noise ratio is many times louder: $300$ for Arvanitaki, $1.4\times10^5$ for Holometer, and $3.3\times10^3$ for Goryachev.
This is consistent with the conclusion drawn from Figure~\ref{fig:h0f}: for our fiducial value of $\kappa$, Advanced LIGO should be able to easily observe orphan memory from high-frequency bursts observed in dedicated high-frequency detectors.


We also investigate the stochastic background from orphan memory.
Stochastic backgrounds arise from the incoherent superposition of many unresolved signals.
They are parameterized by the ratio of energy density in gravitational waves to the total energy density needed to close the Universe~\cite{AllenRomano},
\begin{equation}
\label{eq:Omega}
	\Omega_\text{gw}(f) = \frac{1}{\rho_c}\frac{d\rho_\text{gw}}{d\ln f}
  	= \frac{2\pi^2}{3H_0^2}f^3 S_h(f),
\end{equation}
where $\rho_c$ is the critical energy density of the Universe, $d\rho_\text{gw}$ is the gravitational-wave energy density between $f$ and $f+df$,
$S_h(f)$ is the strain power spectral density of an ensemble of sources, and $H_0$ is the Hubble parameter.

\begin{figure}[h!]
  \centering
    \includegraphics[width=0.5\textwidth]{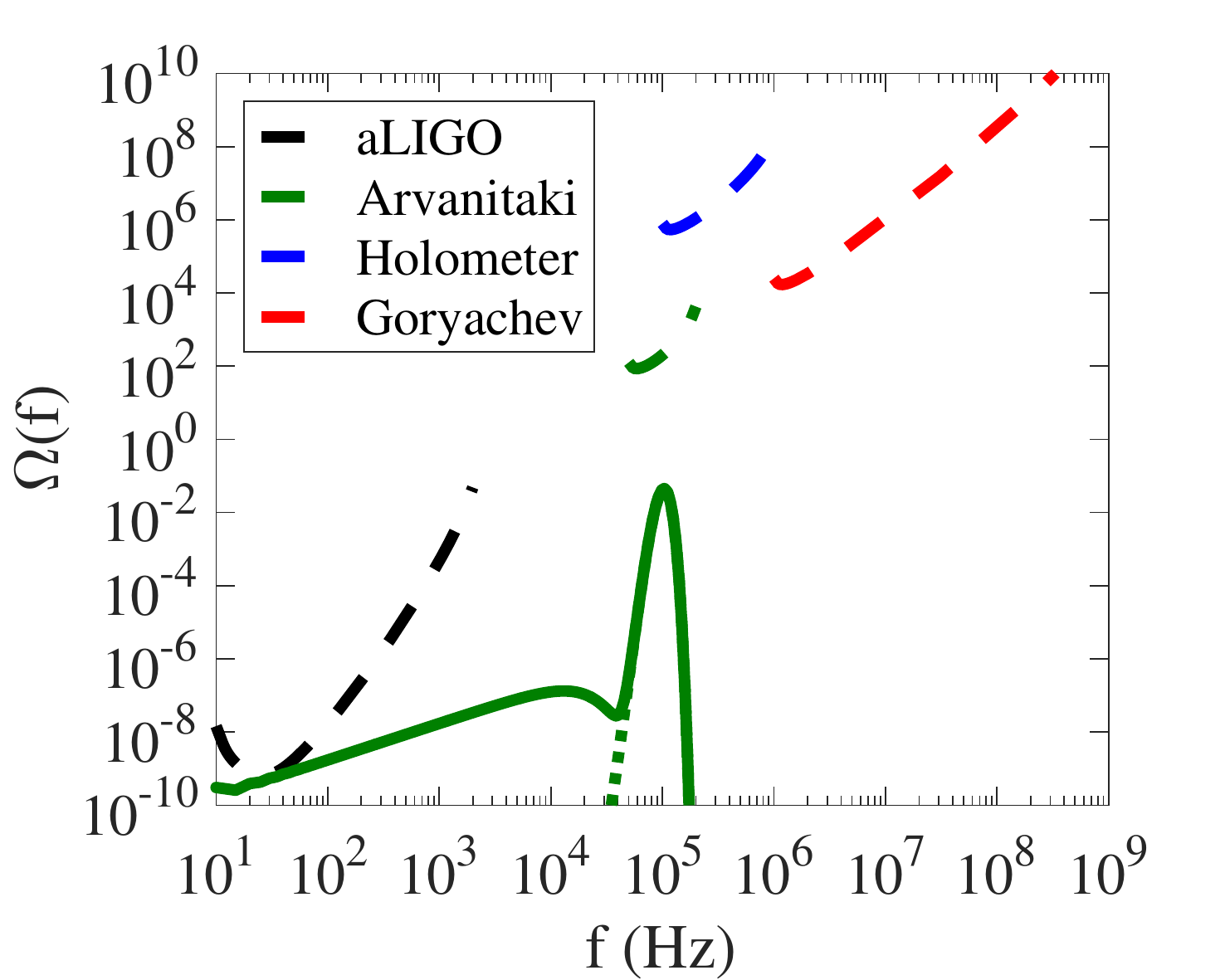}
  \caption{
  	Gravitational-wave energy density spectra.
    Solid green shows the spectra for an ensemble of high-frequency $\unit[100]{kHz}$ bursts with memory.
    The dashed green shows just the oscillatory signal.
    The peak height is chosen so that the signal is just visible by Advanced LIGO after one year of integration.
    The black curve shows the ``power-law integrated curve"~\cite{locus} for Advanced LIGO at design sensitivity.
    A stochastic background spectra that intersects this curve is probably detectable with a cross-correlation signal-to-noise ratio exceeding one.
    Background spectra that do not intersect the power-law integrated curve are not detectable.
    We also include power-law integrated curves for several other high-frequency detectors mentioned in this paper, all calculated assuming a co-located detector pair operated for one year of integration.
For sources with our fiducial value of $\kappa$, it is possible to detect the stochastic background from orphan memory before the signal is observable in dedicated high-frequency detectors.
}
    \label{fig:Omega}
\end{figure}

For an ensemble of sine-Gaussian bursts, $S_h(f)$ is peaked at $f_0$, leading to a peaked distribution of $\Omega_\text{gw}(f)$; see Fig.~\ref{fig:Omega}.
Gravitational-wave memory also creates a stochastic background.
For frequencies $f\ll f_0$,  $S_h(f)\propto f^{-2}$ and so $\Omega_\text{gw}(f)\propto f^{1}$.
The dashed green curve in Fig.~\ref{fig:Omega} shows the stochastic background from an ensemble of $\unit[100]{kHz}$ sine-Gaussian bursts while the solid green curve shows the stochastic background including memory contributions.

In this case, we have tuned the peak height so that the stochastic memory is just barely detectable by Advanced LIGO operating for one year at design sensitivity with cross-correlation signal-to-noise ratio of one~\cite{AllenRomano}.
This is illustrated graphically by the fact that the solid green line intersects the Advanced LIGO ``power-law integrated curve"; see~\cite{locus} for additional details.
We also include power-law integrated curves for the high-frequency detectors included in previous figures.
In each case, we assume a cross-correlation search using a pair of colocated detectors operating for one year.
We find that---given our fiducial choice of $\kappa$---Advanced LIGO is likely to observe an orphan memory background before the peak is observed in high-frequency detectors.

{\it Conclusion.}
In this Letter, we show that orphan memory from high-frequency gravitational-wave sources can be detected when the oscillatory component of the signal is outside of LIGO's frequency band.
Moreover, assuming efficient memory emission, Advanced LIGO is orders of magnitude more sensitive to these bursts than dedicated detectors.
Although no memory-specific LIGO/Virgo searches have been implemented, non-detections by previous ``burst" searches can be converted to limits on the gravitational-wave memory strain,~e.g.,~\cite{LIGO_Burst}.

A dedicated gravitational-wave memory search is desirable.
It will have enhanced sensitivity compared to current burst searches.
Further, a dedicated search can be used to determine whether a detection candidate is consistent with a memory burst by checking to see if the residuals (following signal subtraction) are consistent with Gaussian noise.

High-frequency gravitational-wave sources are conjectural.
It is possible that there are no high-frequency sources in the Universe strong enough to produce orphan memory detectable by Advanced LIGO and successor experiments such as the Einstein Telescope~\cite{EinsteinTelescope} and Cosmic Explorer~\cite{CosmicExplorer}.
However, the generic production of memory by high-frequency sources makes it a useful probe of new physics, which might be revealed by current or planned detectors.
Memory from  lower-frequency (LIGO-band) sources provides a compelling astrophysical target~\cite{Lasky2016}.

\section*{Acknowledgements}
E.T. is supported through ARC FT150100281.
P.D.L. is supported by Australian Research Council DP1410102578 and ARC FT160100112.
This is LIGO document number P1700013.

\bibliography{refs}

\begin{thebibliography}{32}
\expandafter\ifx\csname natexlab\endcsname\relax\def\natexlab#1{#1}\fi
\expandafter\ifx\csname bibnamefont\endcsname\relax
  \def\bibnamefont#1{#1}\fi
\expandafter\ifx\csname bibfnamefont\endcsname\relax
  \def\bibfnamefont#1{#1}\fi
\expandafter\ifx\csname citenamefont\endcsname\relax
  \def\citenamefont#1{#1}\fi
\expandafter\ifx\csname url\endcsname\relax
  \def\url#1{\texttt{#1}}\fi
\expandafter\ifx\csname urlprefix\endcsname\relax\def\urlprefix{URL }\fi
\providecommand{\bibinfo}[2]{#2}
\providecommand{\eprint}[2][]{\url{#2}}

\bibitem[{\citenamefont{{Abbott} et~al.}(2016)\citenamefont{{Abbott}, {Abbott},
  {Abbott} et~al.}}]{GW150914}
\bibinfo{author}{\bibfnamefont{B.~P.} \bibnamefont{{Abbott}}},
  \bibinfo{author}{\bibfnamefont{R.}~\bibnamefont{{Abbott}}},
  \bibinfo{author}{\bibfnamefont{T.~D.} \bibnamefont{{Abbott}}},
  \bibnamefont{et~al.}, \bibinfo{journal}{Phys. Rev. Lett.}
  \textbf{\bibinfo{volume}{116}}, \bibinfo{pages}{061102}
  (\bibinfo{year}{2016}).

\bibitem[{\citenamefont{{Lasky} et~al.}(2016)\citenamefont{{Lasky}, {Thrane},
  {Levin}, {Blackman}, and {Chen}}}]{Lasky2016}
\bibinfo{author}{\bibfnamefont{P.~D.} \bibnamefont{{Lasky}}},
  \bibinfo{author}{\bibfnamefont{E.}~\bibnamefont{{Thrane}}},
  \bibinfo{author}{\bibfnamefont{Y.}~\bibnamefont{{Levin}}},
  \bibinfo{author}{\bibfnamefont{J.}~\bibnamefont{{Blackman}}},
  \bibnamefont{and} \bibinfo{author}{\bibfnamefont{Y.}~\bibnamefont{{Chen}}},
  \bibinfo{journal}{Phys. Rev. Lett.} \textbf{\bibinfo{volume}{117}},
  \bibinfo{pages}{061102} (\bibinfo{year}{2016}).

\bibitem[{\citenamefont{{Zel'dovich} and {Polnarev}}(1974)}]{Zeldovich1974}
\bibinfo{author}{\bibfnamefont{Y.~B.} \bibnamefont{{Zel'dovich}}}
  \bibnamefont{and} \bibinfo{author}{\bibfnamefont{A.~G.}
  \bibnamefont{{Polnarev}}}, \bibinfo{journal}{Sov. Astron.}
  \textbf{\bibinfo{volume}{18}}, \bibinfo{pages}{17} (\bibinfo{year}{1974}).

\bibitem[{\citenamefont{{Braginskii} and {Thorne}}(1987)}]{Braginsky1987}
\bibinfo{author}{\bibfnamefont{V.~B.} \bibnamefont{{Braginskii}}}
  \bibnamefont{and} \bibinfo{author}{\bibfnamefont{K.~S.}
  \bibnamefont{{Thorne}}}, \bibinfo{journal}{Nature}
  \textbf{\bibinfo{volume}{327}}, \bibinfo{pages}{123} (\bibinfo{year}{1987}).

\bibitem[{\citenamefont{{Christodoulou}}(1991)}]{Christ1991}
\bibinfo{author}{\bibfnamefont{D.}~\bibnamefont{{Christodoulou}}},
  \bibinfo{journal}{Phys. Rev. Lett.} \textbf{\bibinfo{volume}{67}},
  \bibinfo{pages}{1486} (\bibinfo{year}{1991}).

\bibitem[{\citenamefont{Favata}(2009)}]{favata09}
\bibinfo{author}{\bibfnamefont{M.}~\bibnamefont{Favata}},
  \bibinfo{journal}{Phys. Rev. D} \textbf{\bibinfo{volume}{80}},
  \bibinfo{pages}{024002} (\bibinfo{year}{2009}).

\bibitem[{\citenamefont{{Favata}}(2010)}]{Favata2009c}
\bibinfo{author}{\bibfnamefont{M.}~\bibnamefont{{Favata}}},
  \bibinfo{journal}{Classical and Quantum Gravity}
  \textbf{\bibinfo{volume}{27}}, \bibinfo{eid}{084036} (\bibinfo{year}{2010}).

\bibitem[{\citenamefont{{Favata}}(2011)}]{favata11}
\bibinfo{author}{\bibfnamefont{M.}~\bibnamefont{{Favata}}},
  \bibinfo{journal}{\prd} \textbf{\bibinfo{volume}{84}},
  \bibinfo{pages}{124013} (\bibinfo{year}{2011}).

\bibitem[{\citenamefont{{Seto}}(2009)}]{Seto2009}
\bibinfo{author}{\bibfnamefont{N.}~\bibnamefont{{Seto}}},
  \bibinfo{journal}{MNRAS} \textbf{\bibinfo{volume}{400}}, \bibinfo{pages}{L38}
  (\bibinfo{year}{2009}).

\bibitem[{\citenamefont{{Pshirkov} et~al.}(2010)\citenamefont{{Pshirkov},
  {Baskaran}, and {Postnov}}}]{Pshirkov10}
\bibinfo{author}{\bibfnamefont{M.~S.} \bibnamefont{{Pshirkov}}},
  \bibinfo{author}{\bibfnamefont{D.}~\bibnamefont{{Baskaran}}},
  \bibnamefont{and} \bibinfo{author}{\bibfnamefont{K.~A.}
  \bibnamefont{{Postnov}}}, \bibinfo{journal}{MNRAS}
  \textbf{\bibinfo{volume}{402}}, \bibinfo{pages}{417} (\bibinfo{year}{2010}).

\bibitem[{\citenamefont{{van Haasteren} and {Levin}}(2010)}]{Vanhaasteren2010}
\bibinfo{author}{\bibfnamefont{R.}~\bibnamefont{{van Haasteren}}}
  \bibnamefont{and} \bibinfo{author}{\bibfnamefont{Y.}~\bibnamefont{{Levin}}},
  \bibinfo{journal}{MNRAS} \textbf{\bibinfo{volume}{401}},
  \bibinfo{pages}{2372} (\bibinfo{year}{2010}).

\bibitem[{\citenamefont{{Arzoumanian} et~al.}(2015)\citenamefont{{Arzoumanian},
  {Brazier}, {Burke-Spolaor} et~al.}}]{Nanograv2015}
\bibinfo{author}{\bibfnamefont{Z.}~\bibnamefont{{Arzoumanian}}},
  \bibinfo{author}{\bibfnamefont{A.}~\bibnamefont{{Brazier}}},
  \bibinfo{author}{\bibfnamefont{S.}~\bibnamefont{{Burke-Spolaor}}},
  \bibnamefont{et~al.}, \bibinfo{journal}{ApJ} \textbf{\bibinfo{volume}{810}},
  \bibinfo{eid}{150} (\bibinfo{year}{2015}).

\bibitem[{\citenamefont{{Wang} et~al.}(2015)\citenamefont{{Wang}, {Hobbs},
  {Coles} et~al.}}]{Wang2015}
\bibinfo{author}{\bibfnamefont{J.~B.} \bibnamefont{{Wang}}},
  \bibinfo{author}{\bibfnamefont{G.}~\bibnamefont{{Hobbs}}},
  \bibinfo{author}{\bibfnamefont{W.}~\bibnamefont{{Coles}}},
  \bibnamefont{et~al.}, \bibinfo{journal}{MNRAS}
  \textbf{\bibinfo{volume}{446}}, \bibinfo{pages}{1657} (\bibinfo{year}{2015}).

\bibitem[{\citenamefont{{Epstein}}(1978)}]{epstein78}
\bibinfo{author}{\bibfnamefont{R.}~\bibnamefont{{Epstein}}},
  \bibinfo{journal}{\apj} \textbf{\bibinfo{volume}{223}}, \bibinfo{pages}{1037}
  (\bibinfo{year}{1978}).

\bibitem[{\citenamefont{{Turner}}(1978)}]{turner78}
\bibinfo{author}{\bibfnamefont{M.~S.} \bibnamefont{{Turner}}},
  \bibinfo{journal}{\nat} \textbf{\bibinfo{volume}{274}}, \bibinfo{pages}{565}
  (\bibinfo{year}{1978}).

\bibitem[{\citenamefont{{Arvanitaki} and {Geraci}}(2013)}]{Arvanitaki}
\bibinfo{author}{\bibfnamefont{A.}~\bibnamefont{{Arvanitaki}}}
  \bibnamefont{and} \bibinfo{author}{\bibfnamefont{A.~A.}
  \bibnamefont{{Geraci}}}, \bibinfo{journal}{Phys. Rev. Lett.}
  \textbf{\bibinfo{volume}{110}}, \bibinfo{pages}{071105}
  (\bibinfo{year}{2013}).

\bibitem[{\citenamefont{{Goryachev} and {Tobar}}(2014)}]{Tobar2014}
\bibinfo{author}{\bibfnamefont{M.}~\bibnamefont{{Goryachev}}} \bibnamefont{and}
  \bibinfo{author}{\bibfnamefont{M.~E.} \bibnamefont{{Tobar}}},
  \bibinfo{journal}{Phys. Rev. D} \textbf{\bibinfo{volume}{90}},
  \bibinfo{eid}{102005} (\bibinfo{year}{2014}).

\bibitem[{\citenamefont{{Chou} et~al.}(2015)\citenamefont{{Chou}, {Gustafson},
  {Hogan}, {Kamai} et~al.}}]{Holometer2015}
\bibinfo{author}{\bibfnamefont{A.~S.} \bibnamefont{{Chou}}},
  \bibinfo{author}{\bibfnamefont{R.}~\bibnamefont{{Gustafson}}},
  \bibinfo{author}{\bibfnamefont{C.}~\bibnamefont{{Hogan}}},
  \bibinfo{author}{\bibfnamefont{B.}~\bibnamefont{{Kamai}}},
  \bibnamefont{et~al.}, \bibinfo{journal}{ArXiv e-prints}
  (\bibinfo{year}{2015}).

\bibitem[{\citenamefont{{Favata}}(2009)}]{Favata2009a}
\bibinfo{author}{\bibfnamefont{M.}~\bibnamefont{{Favata}}},
  \bibinfo{journal}{Apjl} \textbf{\bibinfo{volume}{696}}, \bibinfo{pages}{L159}
  (\bibinfo{year}{2009}).

\bibitem[{noi()}]{noisecurve}
\bibinfo{note}{\url{https://dcc.ligo.org/LIGO-T0900288/public}}.

\bibitem[{\citenamefont{{Cruise}}(2012)}]{Cruise2012}
\bibinfo{author}{\bibfnamefont{A.~M.} \bibnamefont{{Cruise}}},
  \bibinfo{journal}{Classical and Quantum Gravity}
  \textbf{\bibinfo{volume}{29}}, \bibinfo{eid}{095003} (\bibinfo{year}{2012}).

\bibitem[{\citenamefont{{Nakamura} et~al.}(1997)\citenamefont{{Nakamura},
  {Sasaki}, {Tanaka}, and {Thorne}}}]{Nakamura1997}
\bibinfo{author}{\bibfnamefont{T.}~\bibnamefont{{Nakamura}}},
  \bibinfo{author}{\bibfnamefont{M.}~\bibnamefont{{Sasaki}}},
  \bibinfo{author}{\bibfnamefont{T.}~\bibnamefont{{Tanaka}}}, \bibnamefont{and}
  \bibinfo{author}{\bibfnamefont{K.~S.} \bibnamefont{{Thorne}}},
  \bibinfo{journal}{ApJL} \textbf{\bibinfo{volume}{487}}, \bibinfo{pages}{L139}
  (\bibinfo{year}{1997}).

\bibitem[{\citenamefont{{Greene}}(2012)}]{Greene2012}
\bibinfo{author}{\bibfnamefont{J.~E.} \bibnamefont{{Greene}}},
  \bibinfo{journal}{Nature Communications} \textbf{\bibinfo{volume}{3}},
  \bibinfo{eid}{1304} (\bibinfo{year}{2012}), \eprint{1211.7082}.

\bibitem[{\citenamefont{{Kurita} and {Nakano}}(2016)}]{Kurita2015}
\bibinfo{author}{\bibfnamefont{Y.}~\bibnamefont{{Kurita}}} \bibnamefont{and}
  \bibinfo{author}{\bibfnamefont{H.}~\bibnamefont{{Nakano}}},
  \bibinfo{journal}{Phys. Rev. D} \textbf{\bibinfo{volume}{93}},
  \bibinfo{eid}{023508} (\bibinfo{year}{2016}).

\bibitem[{\citenamefont{{Damour} and {Vilenkin}}(2000)}]{Damour1}
\bibinfo{author}{\bibfnamefont{T.}~\bibnamefont{{Damour}}} \bibnamefont{and}
  \bibinfo{author}{\bibfnamefont{A.}~\bibnamefont{{Vilenkin}}},
  \bibinfo{journal}{Phys. Rev. Lett.} \textbf{\bibinfo{volume}{85}},
  \bibinfo{pages}{3761} (\bibinfo{year}{2000}).

\bibitem[{\citenamefont{{Seahra} et~al.}(2005)\citenamefont{{Seahra},
  {Clarkson}, and {Maartens}}}]{Seahra2005}
\bibinfo{author}{\bibfnamefont{S.~S.} \bibnamefont{{Seahra}}},
  \bibinfo{author}{\bibfnamefont{C.}~\bibnamefont{{Clarkson}}},
  \bibnamefont{and}
  \bibinfo{author}{\bibfnamefont{R.}~\bibnamefont{{Maartens}}},
  \bibinfo{journal}{Phys. Rev. Lett.} \textbf{\bibinfo{volume}{94}},
  \bibinfo{eid}{121302} (\bibinfo{year}{2005}).

\bibitem[{\citenamefont{{Clarkson} and {Seahra}}(2007)}]{Clarkson2007}
\bibinfo{author}{\bibfnamefont{C.}~\bibnamefont{{Clarkson}}} \bibnamefont{and}
  \bibinfo{author}{\bibfnamefont{S.~S.} \bibnamefont{{Seahra}}},
  \bibinfo{journal}{Classical and Quantum Gravity}
  \textbf{\bibinfo{volume}{24}}, \bibinfo{pages}{F33} (\bibinfo{year}{2007}).

\bibitem[{\citenamefont{Allen and Romano}(1999)}]{AllenRomano}
\bibinfo{author}{\bibfnamefont{B.}~\bibnamefont{Allen}} \bibnamefont{and}
  \bibinfo{author}{\bibfnamefont{J.~D.} \bibnamefont{Romano}},
  \bibinfo{journal}{Phys. Rev. D} \textbf{\bibinfo{volume}{59}},
  \bibinfo{pages}{102001} (\bibinfo{year}{1999}).

\bibitem[{\citenamefont{Thrane and Romano}(2013)}]{locus}
\bibinfo{author}{\bibfnamefont{E.}~\bibnamefont{Thrane}} \bibnamefont{and}
  \bibinfo{author}{\bibfnamefont{J.~D.} \bibnamefont{Romano}},
  \bibinfo{journal}{Phys. Rev. D} \textbf{\bibinfo{volume}{88}},
  \bibinfo{pages}{124032} (\bibinfo{year}{2013}).

\bibitem[{\citenamefont{Abadie et~al.}(2012)}]{LIGO_Burst}
\bibinfo{author}{\bibfnamefont{J.}~\bibnamefont{Abadie}} \bibnamefont{et~al.},
  \bibinfo{journal}{Phys. Rev. D} \textbf{\bibinfo{volume}{85}},
  \bibinfo{pages}{122007} (\bibinfo{year}{2012}).

\bibitem[{\citenamefont{Punturo et~al.}(2010)}]{EinsteinTelescope}
\bibinfo{author}{\bibfnamefont{M.}~\bibnamefont{Punturo}} \bibnamefont{et~al.},
  \bibinfo{journal}{Classical Quantum Gravity} \textbf{\bibinfo{volume}{27}},
  \bibinfo{pages}{194002} (\bibinfo{year}{2010}).

\bibitem[{\citenamefont{Abbott et~al.}(2016)}]{CosmicExplorer}
\bibinfo{author}{\bibfnamefont{B.~P.} \bibnamefont{Abbott}}
  \bibnamefont{et~al.} (\bibinfo{year}{2016}),
  \bibinfo{note}{https://arxiv.org/abs/1607.08697}.

\end{thebibliography}

\label{lastpage}

\end{document}